\documentclass[a4paper,11pt]{article}

\usepackage{fullpage,amsthm,verbatim,color,xcolor}
\usepackage{ogonek}

\definecolor{myurlcolor}{rgb}{0,0,0.4}
\definecolor{mycitecolor}{rgb}{0,0.5,0}
\definecolor{myrefcolor}{rgb}{0.5,0,0}
\usepackage{hyperref}
\hypersetup{colorlinks,
linkcolor=myrefcolor,
citecolor=mycitecolor,
urlcolor=myurlcolor}

\usepackage[draft]{fixme}
\usepackage{multicol}
\usepackage{changes}

\usepackage{amsmath,bbm}
\usepackage{graphicx}
\usepackage{amsfonts}
\usepackage{amssymb}

\newcommand{\one}{\leavevmode\hbox{\small1\normalsize\kern-.33em1}}

\def\be{\begin{equation}}
\def\ee{\end{equation}}
\def\ben{\begin{eqnarray}}
\def\een{\end{eqnarray}}
\def\eea{\end{array}}
\def\bea{\begin{array}}

\newcommand{\bei}{\begin{itemize}}
\newcommand{\eei}{\end{itemize}}
\newcommand{\ket}[1]{|#1\rangle}
\newcommand{\bra}[1]{\langle#1|}

\newcommand{\eg}{{\it{e.g.~}}}

\newcommand{\etal}{{\it{et al.~}}}

\newcommand{\figref}[1]{Fig.~\ref{#1}}

\theoremstyle{plain}

\theoremstyle{definition}

\theoremstyle{remark}

\begin{document}

\title{Local orthogonality as a multipartite principle for quantum correlations}
\author{ T. Fritz$^{1,2}$, A.~B. Sainz$^{1}$, R. Augusiak$^{1}$, J. Bohr Brask$^{1}$,\\ R.
Chaves$^{1,3}$, A. Leverrier$^{1,4,5}$, and A. Ac\'in$^{1,6}$\\[0.5em]
{\it\small $^1$ICFO--Institut de Ciencies Fotoniques, E--08860 Castelldefels, Barcelona, Spain}\\
{\it\small $^2$Perimeter Institute for Theoretical Physics, Waterloo, Ontario, Canada}\\
{\it\small $^3$Institute for Physics, University of Freiburg, D-79104 Freiburg, Germany}\\
{\it\small $^4$Institute for Theoretical Physics, ETH Z{\"u}rich, 8093 Z{\"u}rich, Switzerland}\\
{\it\small $^5$INRIA Rocquencourt, Domaine de Voluceau, B.P. 105, 78153 Le Chesnay Cedex, France}\\
{\it\small $^6$ICREA--Institucio Catalana de Recerca i Estudis
Avan\c{c}ats, E--08010 Barcelona, Spain}}
\date{\small Journal of Reference: Nature Communications 4, 2263 (2013).  http://dx.doi.org/10.1038/ncomms3263}

\maketitle

\begin{abstract}
In recent years, the use of information principles to understand
quantum correlations has been very successful. Unfortunately, all
principles considered so far have a bipartite formulation, but
intrinsically multipartite principles, yet to be discovered, are
necessary for reproducing quantum correlations. Here, we introduce
local orthogonality, an intrinsically
multipartite principle stating that events involving different
outcomes of the same local measurement must be exclusive, or
orthogonal. We prove that it is
equivalent to no-signaling in the bipartite scenario but more
restrictive for more than two parties.  By exploiting this non-equivalence, it is then demonstrated that some bipartite
supra-quantum
correlations do violate local orthogonality when distributed among
several parties. Finally, we show how its multipartite character
allows revealing the non-quantumness of correlations for which any
bipartite principle fails. We believe that local orthogonality is
a crucial ingredient for understanding no-signaling and quantum
correlations.
\end{abstract}

Understanding the structure of correlations within our current
description of nature, based on quantum physics, is a fundamental
open problem. In particular, one would like to characterize the
set of quantum correlations, i.e.~correlations which can result
from local measurements on quantum states. Pioneering
work by Popescu and Rohrlich showed that the no-signaling
principle -- that is, the impossibility of instantaneous
communication -- does not suffice to recover this quantum
set~\cite{pr}. Indeed, they provided examples of correlations
between two parties compatible with the no-signaling principle but
without any quantum realization. The most paradigmatic example of
these supra-quantum correlations is the so-called Popescu-Rohrlich
(PR) box, also studied by Tsirelson~\cite{tsirelson}. The search
for better principles separating supra-quantum correlations from
quantum ones, or ideally a complete characterization of
the quantum set, was started.

An important boost to this search was due to Van Dam, who
introduced the idea that the existence of supra-quantum
correlations, while not violating the no-signaling principle,
could have implausible consequences from an information processing
point of view. Van Dam showed that distant parties having access
to PR-boxes can render communication complexity trivial and
argued that this could be a reason for the non-existence of these
correlations in Nature~\cite{vd}. Since then, intensive effort has
been devoted to the search for information principles
characterizing the set of quantum correlations, \eg~the
aforementioned non-trivial communication
complexity~\cite{vd,Brassard2006}, information
causality~\cite{Pawlowski2009a}, and macroscopic
locality~\cite{nw}.

Recently, it has been shown that intrinsically multipartite
principles are essential to characterize the set of quantum
correlations. It was proven in~\cite{gallego} that there exist
supra-quantum correlations for three parties that cannot be
detected by any bipartite principle. Unfortunately, most of the
existing principles for quantum correlations are formulated in a
bipartite setting and their multipartite generalization is
unclear, apart from the trivial one in which the parties are split
in two groups and the principle is applied to each bipartition. In
this sense, note that even the no-signaling principle has a
bipartite formulation in the multipartite scenario: correlations
among $n$ distant parties satisfy the no-signaling principle
whenever the marginal distribution seen by a subset of the $n$
parties  is independent of the choice of measurements by the
remaining ones.

In this work, we introduce the concept of local orthogonality
(LO), an intrinsically multipartite principle for correlations.
The principle is based on a definition of orthogonality (or
exclusiveness) between events involving measurement choices and
results by $n$ distant parties: we define some events to be
orthogonal, or exclusive, whenever they involve different outcomes
of the same local measurement by at least one of the parties.
Operationally, we demand that the sum of the probabilities of
mutually exclusive events is less than or equal to one, which
implies a restriction on possible correlations. We provide an
information processing interpretation of this new principle in
terms of a distributed guessing problem.  We then show how the
principle implies a highly non-trivial structure in the space of
correlations. First, we prove that, while the set of LO
correlations coincides with the set of no-signaling  (NS)
correlations for two parties, it is strictly smaller for more than
two parties.  Second, by exploiting the non-equivalence of LO and
the no-signaling principle in the multipartite case, it is
demonstrated how LO can also be used to detect the non-quantumness
of some supra-quantum, bipartite NS correlations. In particular,
we prove that distributed copies of the PR-box violate LO, and
show how one can get very close to the boundary of the quantum set
of bipartite correlations (Tsirelson's bound) by imposing LO.
Finally, we prove that the intrinsically multipartite formulation
of the principle allows one to detect supra-quantum correlations
for which any bipartite principle fails.

In deriving all these results, we exploit a connection between LO
correlations and graph theory, related to the constructions
derived in~\cite{csw} in the study of quantum contextuality.
 In fact, there is a natural notion of orthogonality (also called
"exclusiveness") between events in contextuality scenarios, which are
defined by sets of measurements
sharing some measurement outcomes (two events are orthogonal
if they correspond to different outcomes of a given measurement).
 For instance, several works
have studied how the violation of the corresponding orthogonality
conditions can lead to supra-quantum
correlations~\cite{specker, spekkens}, or how these
orthogonality conditions can be used to provide upper bounds to
the quantum violation of non-contextuality
inequalities~\cite{csw}.

\section*{Results}

\textbf{Definition.} We start by presenting
the formal definition of the LO principle in a general Bell
scenario involving $n$ distant parties, each of them having access
to a physical system. Each party can perform $m$ different
measurements on the system, getting one out of $d$ possible
outcomes. This scenario is denoted by $(n,m,d)$. The measurement
applied by party $i$ is denoted by $x_i$, and the corresponding
outcome by $a_i$, with $i \in \{1,\ldots,n\}$, $x_i\in
\left\{0,\ldots,m-1\right\}$, and $a_i\in \{0,\ldots,d-1\}$. The
correlations among the parties are described by the joint
conditional probability distribution
$P\,(a_1\ldots a_n|x_1\ldots x_n)$, representing the probability
for the parties to get outcomes $a_1,\ldots,a_n$ when making
measurements $x_1,\ldots,x_n$.

The main aim of LO is to introduce a notion of exclusiveness in
the space of events. Consider first two different events $e$ and
$e'$ given by $e=(a_1 \ldots a_n|x_1 \ldots x_n)$ and
$e'=(a'_1\ldots a'_n|x'_1\ldots x'_n)$. We call these two events
locally orthogonal or simply orthogonal, if they involve different
outputs of the same measurement by (at least) one party. That is,
if for some $i$ we have $a_i \neq a_i'$ while $x_i=x_i'$. We then
call a collection of events $\{e_i\}$ orthogonal, or exclusive, if
the events are pairwise orthogonal, and impose that for any set of
orthogonal events, the sum of their probabilities must not be
larger than one,
\begin{equation}
\label{loineq}
\sum_{i} P\,(e_i)\leq 1.
\end{equation}
This requirement is the LO principle.  To summarize: the LO
principle (i) introduces a notion of orthogonality between two
events, (ii) imposes that any number of events are orthogonal
whenever they are pairwise orthogonal, and (iii) requires that the
inequality~(\ref{loineq}) is satisfied for any set of orthogonal
events.

The notion of orthogonality is rather natural in the case of two
events, and the requirement~(\ref{loineq}) is automatically
satisfied for NS correlations. In fact, consider two LO events
$e_1$, $e_2$ with $a_i \neq a_i'$ and $x_i=x_i'$ as before. These
two events can be seen as different outcomes of a correlated
measurement in which: (i) party $i$ first measures $x_i$ and
announces the outcome to the other parties and (ii) the other
parties apply measurements depending on this outcome, in
particular they measure $x_1,\ldots,x_n$, if the outcome is $a_i$,
and $x'_1,\ldots,x'_n$ otherwise.  Thus, as $e_1$ and $e_2$
are outcomes of the same (correlated) measurement, the
requirement~(\ref{loineq}) immediately follows from normalization.
Note that the no-signaling principle is essential for this
correlated measurement to be meaningful, as it is possible to
define the marginal probability for party $i$ in the first step of
the correlated measurement independently of the successive actions
by the other parties.

 The principle however becomes more restrictive when
considering more events, where the previous reasoning does not
apply any longer. As mentioned above, we extend the initial
definition of orthogonality for two events to more events by
demanding pairwise orthogonality. It is precisely this extended
and intrinsically multipartite formulation that makes the
principle non-trivial, because it involves summing probabilities
conditioned on different measurements. Taken together, all the
restrictions on the conditional probability distributions implied
by LO define the set of LO inequalities~(\ref{loineq}) for the
$(n,m,d)$ scenario. The set of LO correlations is then the set of
conditional distributions $P\,(a_1 \ldots a_n|x_1 \ldots x_n)$
satisfying all the LO inequalities.

Moreover, we can define the sets of conditional probability distributions
that obey LO for a certain number of copies in the following
sense. A given distribution for the $(n,m,d)$ scenario can be
thought of as provided by some device shared between the $n$
parties each having access to one input and output of the device.
If the distribution provided by the device is compatible with LO,
a natural question is whether a larger distribution coming from
several copies of such a device distributed among more parties
necessarily satisfies LO. As we will explain below, the answer to
this question is negative. That is, LO displays activation
effects and, hence, we have a hierarchy of sets. The largest set
in this hierarchy, denoted $\mathcal{LO}^1$, is the set of all distributions in the
$(n,m,d)$ scenario which obey the LO inequalities for this
scenario. Now consider $k$ copies of a device characterized by
a distribution $P$, distributed among $kn$ parties, each of which
has access to one input of only one device. If the
distribution $P^k$ of the $kn$-partite global device obeys all the
LO inequalities for the scenario $(kn,m,d)$, we say that $P$
satisfies LO$^k$, and belongs to $\mathcal{LO}^k$. We denote by
$\mathcal{LO}^\infty$ the set of distributions for $(n,m,d)$ which
obey the LO inequalities for any number of copies.

\begin{figure}
\centering
\includegraphics[scale=0.20]{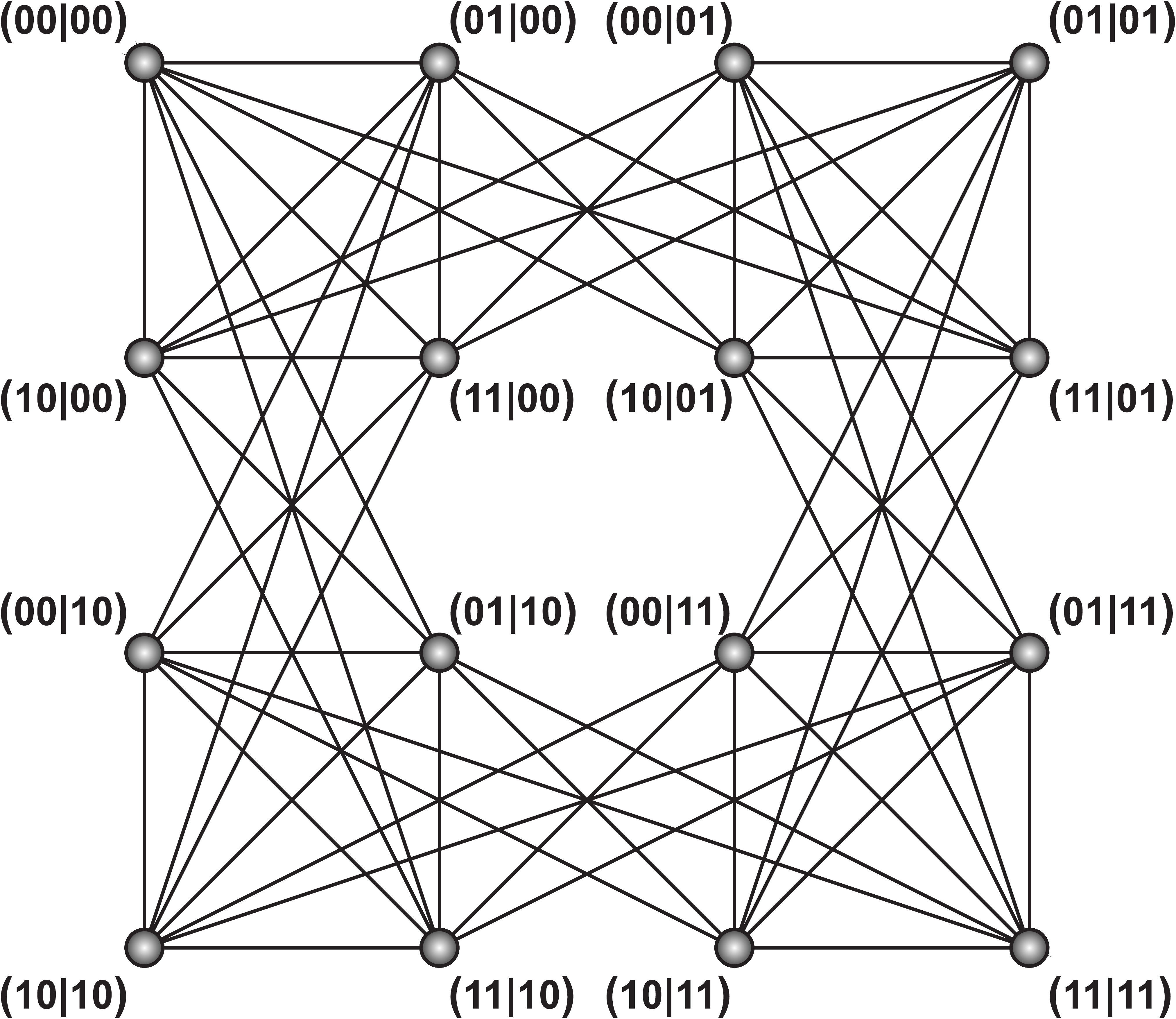}
\caption{Orthogonality graph of the $(2,2,2)$ scenario. As
mentioned in the text, each possible event corresponds to a node,
while the edges connect locally orthogonal events.} \label{OG2}
\end{figure}

Having stated the LO principle, our main goal is the study of the
sets $\mathcal{LO}^k$ of LO correlations. As we shall see, the LO
principle turns out to be very powerful for ruling out non-quantum
correlations. As in the case of contextuality~\cite{csw}, graph
theory is perfectly suited for our purposes. We consider the
$(md)^n$ possible events in the $(n,m,d)$ scenario and map them
onto an orthogonality graph with $(md)^n$ vertices, where two
vertices are connected by an edge if and only if the corresponding
events are locally orthogonal. For instance, \figref{OG2} shows
the orthogonality graph of the $(2,2,2)$ scenario. In graph
theory, a clique in a graph $G=(V,E)$ is a subset of vertices $C
\subseteq V$ such that the subgraph induced by $C$ is complete,
i.e.~such that all pairs of vertices in $C$ are connected by an
edge in $G$. A clique is maximal if it cannot be extended to
another clique by including a new vertex. Clearly, any clique in
the orthogonality graph of events gives rise to an LO inequality
(and vice versa), as all events in the clique are connected and,
thus, are pairwise orthogonal. Therefore, the problem ``find all
the optimal LO inequalities'' is equivalent to ``find all maximal
cliques of the associated orthogonality graph''. While the problem
of finding all maximal cliques of a graph is known to be
NP-hard~\cite{clique_np-hard}, there exist software packages
\cite{mace,cliquer} that provide the solution for small graphs. We
have used these packages to derive and partly classify LO
inequalities for various Bell scenarios, which will be presented
elsewhere. Note however that in principle, while the problem of
finding the maximal cliques is NP-hard for general graphs, this
may no longer  be the case for graphs associated to correlations
among distant parties. Indeed, these graphs may represent a subset
of all possible graphs
that does not include the hard instances of the problem.\\

\noindent\textbf{Distributed guessing
problems.} Before moving on to the characterization of LO
correlations, we provide an interpretation of the principle from
an information processing viewpoint. To this end, we introduce the
notion of a Distributed Guessing Problem (DGP).

\begin{figure}
\centering
\includegraphics[width=0.5\linewidth]{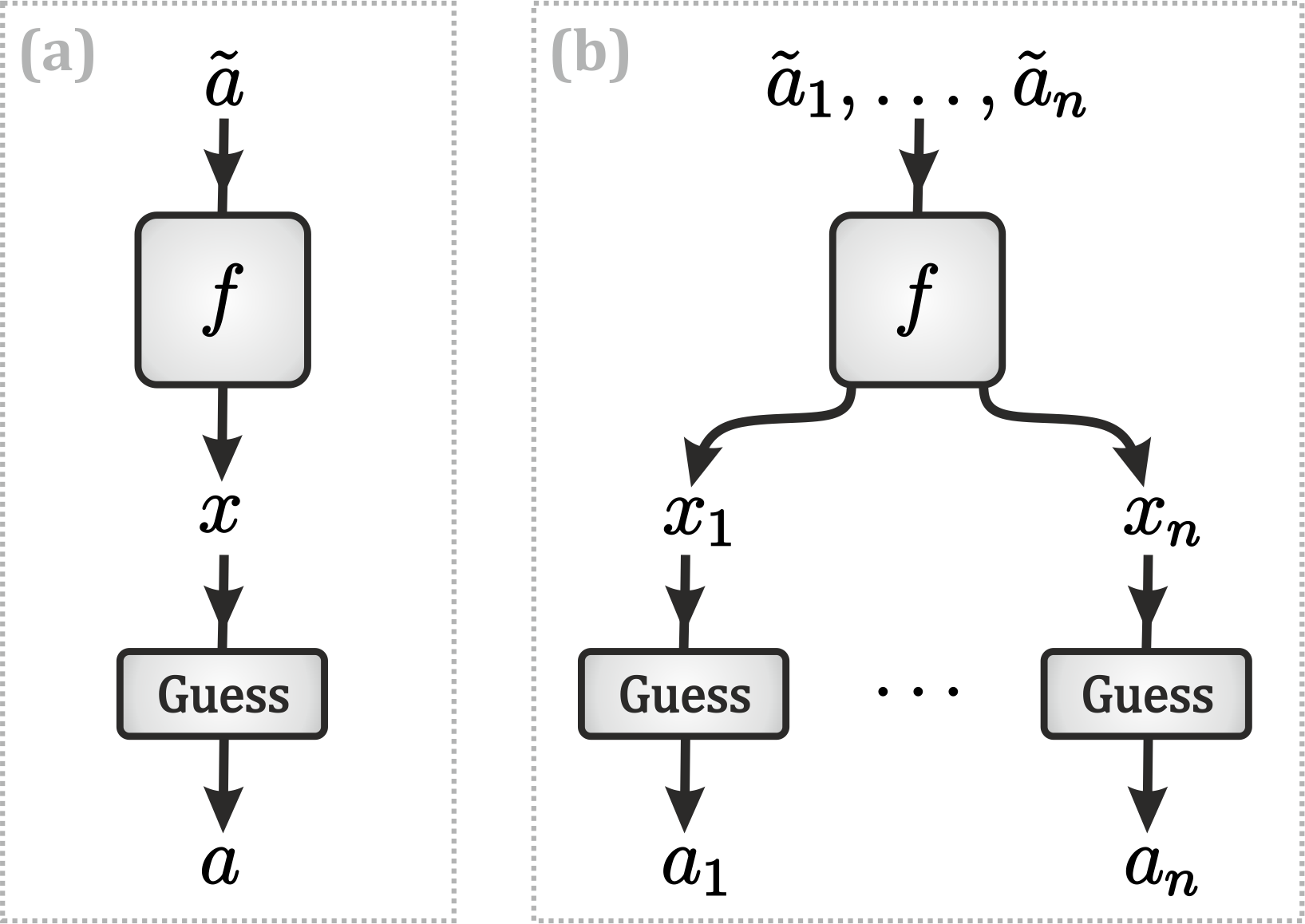}
\caption{Distributed guessing problems. In the standard scenario (a), an observer has to guess the
value of a parameter $\widetilde a$ given only some function of
it, $x=f(\widetilde a)$. In the distributed scenario (b), the input
parameter is a vector of $n$ symbols, $(\widetilde
a_1,\ldots,\widetilde a_n)$ and so is the data given to the
players $(x_1,\ldots,x_n)$. Each of player has access to just one
of the symbols $x_j$ and has to guess the corresponding initial
parameter $\widetilde a_j$. The game is won when all players guess
correctly.} \label{dgp}
\end{figure}

Guessing problems are ubiquitous in science. In the standard
formulation (see \figref{dgp}(a)) an observer has access to some
data $x$ which depends on some initial parameter $\widetilde a$,
that is $x=f(\widetilde a)$. From the observed data, the observer
should make a guess, $a$, of the initial parameter. His goal is to
maximize the probability of guessing correctly.

Guessing problems can be easily adapted to distributed scenarios.
It is convenient to present the distributed guessing problem as a
game and to phrase it in terms of vectors of symbols. Consider
then a non-local game in which a referee has access to a set of
vectors of $n$ symbols with values in $\{0,\ldots,d-1\}$. Denote
this set by $S$ and by $|S|$ its size, which can be less than
$d^n$ in general. Now, the referee chooses a vector
$(\widetilde{a}_1,\ldots,\widetilde{a}_n)$ uniformly at random
from $S$, and encodes it into a new vector of, again, $n$ symbols
using a function $f$. However, the new symbols can now take $m$
values and, thus, $f : S \longrightarrow \{0,\ldots,m-1\}^n$. The
resulting vector is $(x_1,\ldots,x_n) = f(\widetilde
a_1,\ldots,\widetilde a_n)$. These $n$ symbols are distributed
among $n$ distant players who cannot communicate and must produce
individual guesses $a_1,\ldots,a_n$. Their goal is to guess the
initial input to the function, that is, they win whenever
$a_j=\widetilde{a}_j$ for all $j$. Note that the encoding function
$f$ and the set $S$ are known in advance to all the players.

For some $f$, e.g.~for $x_j = \widetilde a_j$, this task is very
simple. However, there exist functions for which the task becomes
extremely difficult. For a fixed size $|S|$, the most difficult
functions are those for which the maximum guessing probability is
equal to $1/|S|$. The players can always achieve this guessing
probability by agreeing beforehand on one of the $|S|$ possible
outputs, which they output regardless of the $x_j$. Since the
input is uniform on $S$, their guess is correct with probability
$1/|S|$. A DGP is thus  maximally difficult whenever this strategy
is optimal, that is, whenever it is impossible to provide a better
estimate of the input than random guessing. For such an $f$,
having access to the symbols $x_j$ does not provide any useful
information to the parties. Note that non-trivial maximally
difficult functions are possible only in distributed scenarios. In
standard single-observer guessing problems, the only maximally
difficult function is the one defined by a constant function $f$,
which trivially erases any information about the input. An example
of a difficult function in the $(n,2,2)$ scenario for odd $n$ and
classically correlated players is $f(a_1,\ldots,a_n) =(a_n, a_1,
\ldots, a_{n-1})$ defined on the set $S$ of inputs satisfying $
a_1\oplus\ldots\oplus a_n = 0$. This is the
guess-your-neighbor's-input task considered in~\cite{gyni}.

 As we prove in Methods, a DGP is maximally difficult for players sharing classical correlations (classical players) if, and only if, it corresponds to an LO inequality. Hence, in order to win the game with a probability larger than $1/|S|$, they need to share correlations violating LO. In particular, quantum correlations provide no advantage over the trivial strategy of randomly guessing the solution.\\

\noindent\textbf{The no-signaling principle.} The first question
we ask when characterizing LO correlations is how they relate to
the set of no-signaling correlations, denoted by $\mathcal{NS}$.
For bipartite scenarios, the two principles define the same set of
correlations, as was already noticed in~\cite{csw} (in Methods, we
give a slightly different proof that emphasizes the connection
with LO). However, the equivalence between LO and NS breaks down
when moving to the multipartite scenario. We exploited the
graph-theoretical approach mentioned in the previous section to
generate the list of LO inequalities for different scenarios, and
then classified them into equivalence classes under relabellings,
permutations of parties, and no-signaling constraints (using a
Mathematica as well as MATLAB code kindly provided to us by
J.~D.~Bancal). Already in the simplest tripartite scenario
$(3,2,2)$, we find one and only one class of non-trivial LO
inequalities, where non-trivial means that the inequalities are
violated by some NS correlations. This inequality turns out to be
the GYNI inequality~\cite{gyni}, which in the tripartite case
reads $P(000|000)+P(110|011)+P(011|101)+P(101|110) \leq 1$. It is
easy to see by simple inspection that GYNI is an LO inequality. As
shown in~\cite{gyni}, the maximum of the GYNI inequality over
$\mathcal{NS}$ is equal to $4/3$. Our numerical data suggest that
the gap between $\mathcal{LO}^1$ and $\mathcal{NS}$ increases with
the number of parties: in the $(4,2,2)$ scenario, we find 35
equivalence classes. Unfortunately, for more parties ($n>4$), even
the simplest scenario $(n,2,2)$ becomes computationally
intractable due to the large size of the orthogonality graph.
Nevertheless, examples of such inequalities for larger $n$ as well
as $m$ and $d$ are known and can be constructed from unextendible
product bases~\cite{BennettUPB} by using the method discussed
in~\cite{augusiak2011,augusiak2012} (see also \cite{bookchapter}).

 It is worth mentioning that all the known examples of
non-trivial, tight  (in the sense of defining a tight Bell
inequality \cite{Mas03}) information tasks with no quantum advantage, given in
Refs.~\cite{gyni,augusiak2011,augusiak2012,bookchapter}, are
examples of LO inequalities. It is an interesting working
conjecture to prove that  any non-trivial and tight
information task with no quantum advantage defines an LO
inequality. In particular, this would
imply that any  non-trivial tight Bell inequality in a bipartite scenario has quantum violations.\\

\noindent\textbf{Supraquantum
correlations.} The LO principle is naturally
satisfied by quantum correlations. Indeed, orthogonal events can
be associated with measurements described by projectors with
disjoint supports (see Methods for details). We now investigate
the use of the LO principle as a tool to detect post-quantum
no-signaling correlations. Clearly, those correlations violating
GYNI are in contradiction with LO as well. However, the situation
turns out to be much richer, already for two parties. In principle,
one might think that LO would be useless for the detection of
supra-quantum bipartite correlations because of the equivalence
with NS. However, this intuition is not correct. We can show that postulating LO also on the many-party level
leads to detection of non-quantumness of bipartite
correlations. Given some bipartite correlations, the main idea
consists in distributing $k$ copies of these among $2k$ parties,
such that one party has access to one part  (input and output) of only one bipartite
box. In the resulting $2k$-partite scenario, the LO principle is
stronger than the NS principle. Thus, it may happen that the
initial bipartite correlations violate LO when distributed among
different parties in a network. The idea is similar in spirit to
the network approach to non-locality presented in
Ref.~\cite{natcomm}.

A PR-box is a hypothetical device taking binary inputs and giving
binary outputs which obey $PR(ab|xy)= 1/2$ if $a \oplus b =xy$ and
$PR(ab|xy)= 0$ otherwise~\cite{pr}. These boxes are known to be
more non-local than what quantum theory allows. For instance, they
provide a violation of the Clauser-Horne-Shimony-Holt Bell
inequality~\cite{chsh} larger than Tsirelson's bound for quantum
correlations~\cite{tsirelson2}. PR-boxes are bipartite
no-signaling devices, and therefore might na\"ively be expected to
satisfy LO. However, we prove now that when distributed in
networks they violate LO.
\begin{figure}
\centering
\includegraphics[width=0.5\linewidth]{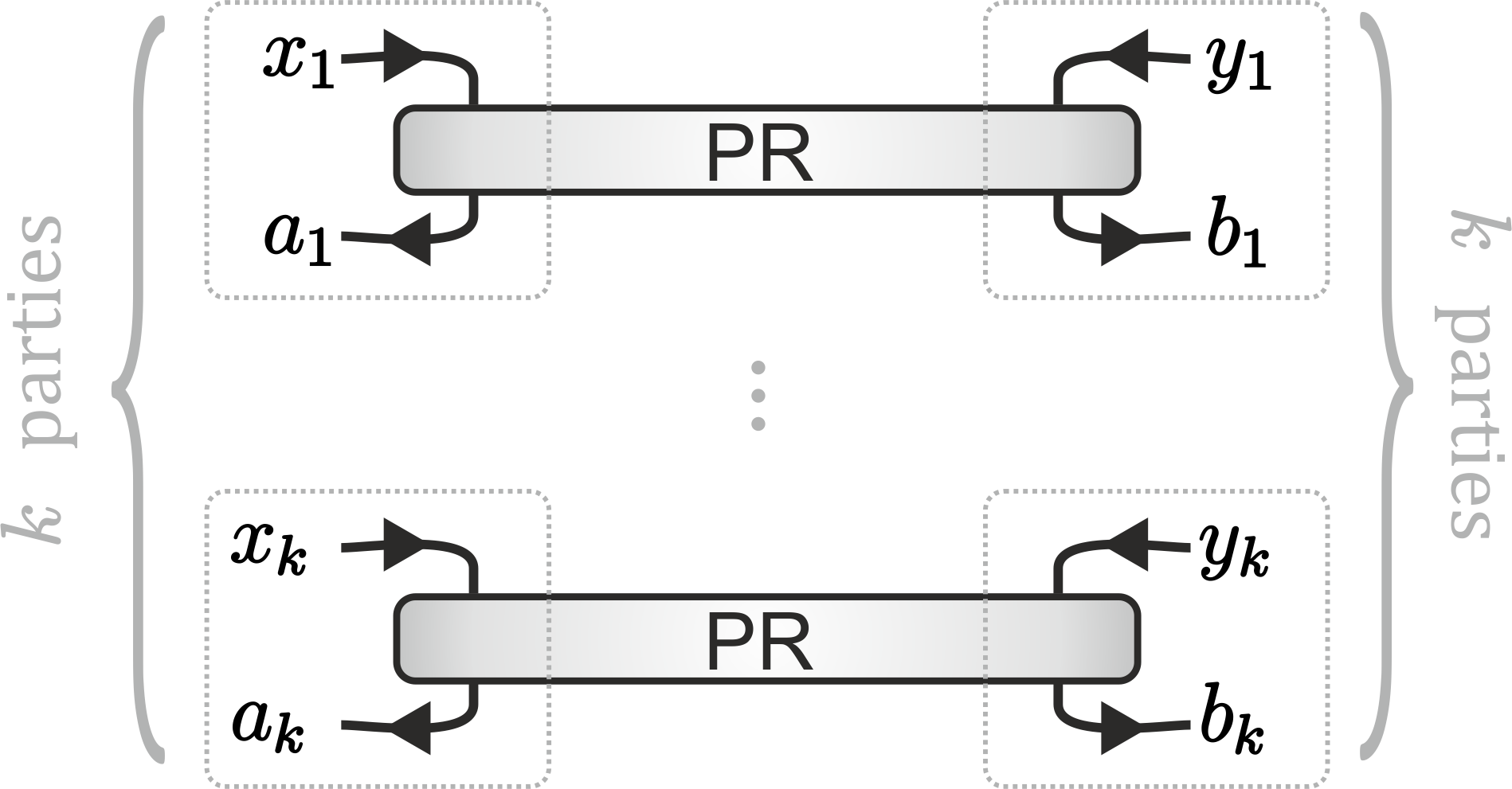}
\caption{Network of PR-boxes. We consider $k$ copies of a PR-box
shared among $2k$ parties. Each party has access to one part of a
box.} \label{f-manyPR}
\end{figure}
Consider $k$ copies of a PR-box, distributed among $2k$ parties as shown in \figref{f-manyPR}. The
conditional probability distribution is:
\begin{equation}
\label{def-NPR}
P(a_1 b_1 \cdots a_k b_k | x_1y_1 \cdots x_k y_k ) = \prod_{j=1}^k PR(a_j b_j | x_j y_j) \,,
\end{equation}
where $j$ labels the $k$ boxes. Already for $k=2$, we find LO
inequalities violated by these two copies of the PR-box, and hence
PR boxes do not satisfy LO$^2$. One example of such an inequality
is $
P(0000|0000)+P(1110|0011)+P(0011|0110)+P(1101|1011)+P(0111|1101)\leq
1.$ For a PR-box, the left-hand side is equal to $5/4$. We can
also analyze noisy versions of the PR-box given by $P_{q} = q \,PR
+ (1-q) P_{\mathbb{I}}$, where $P_\mathbb{I} (a b|x y)=1/4$ for
all $a,b,x,y$. We find that two copies of a noisy PR-box violate
LO down to $q\approx 0.72$, which is close to Tsirelson's bound
$q=1/\sqrt{2}\approx 0.707$ (meaning that noisy boxes with $q \leq
1/\sqrt{2}$ can be simulated with quantum states and
measurements). Detailed derivations of these results are presented
in Methods.

An immediate consequence of these results is that LO also rules
out all extremal boxes in the $(2,2,d)$ and $(2,m,2)$ scenarios.
The first case follows from the fact that any extremal box in the
$(2,2,d)$ case can always be used to simulate a PR-box arbitrarily
well~\cite{BarrettPRA2005}. For the $(2,m,2)$ scenario, all the
extremal boxes were characterized in~\cite{JM05} (see also
\cite{BP05}). There, it is shown that, up to symmetries, an
extremal nonlocal box is equivalent to a PR-box if one restricts
the considerations to the first two inputs out of the $m$ possible
choices. Hence, any LO inequality violated by a PR-box, is also
violated by any extremal non-local bipartite binary box.

\begin{figure}
\centering
\includegraphics[scale=0.55]{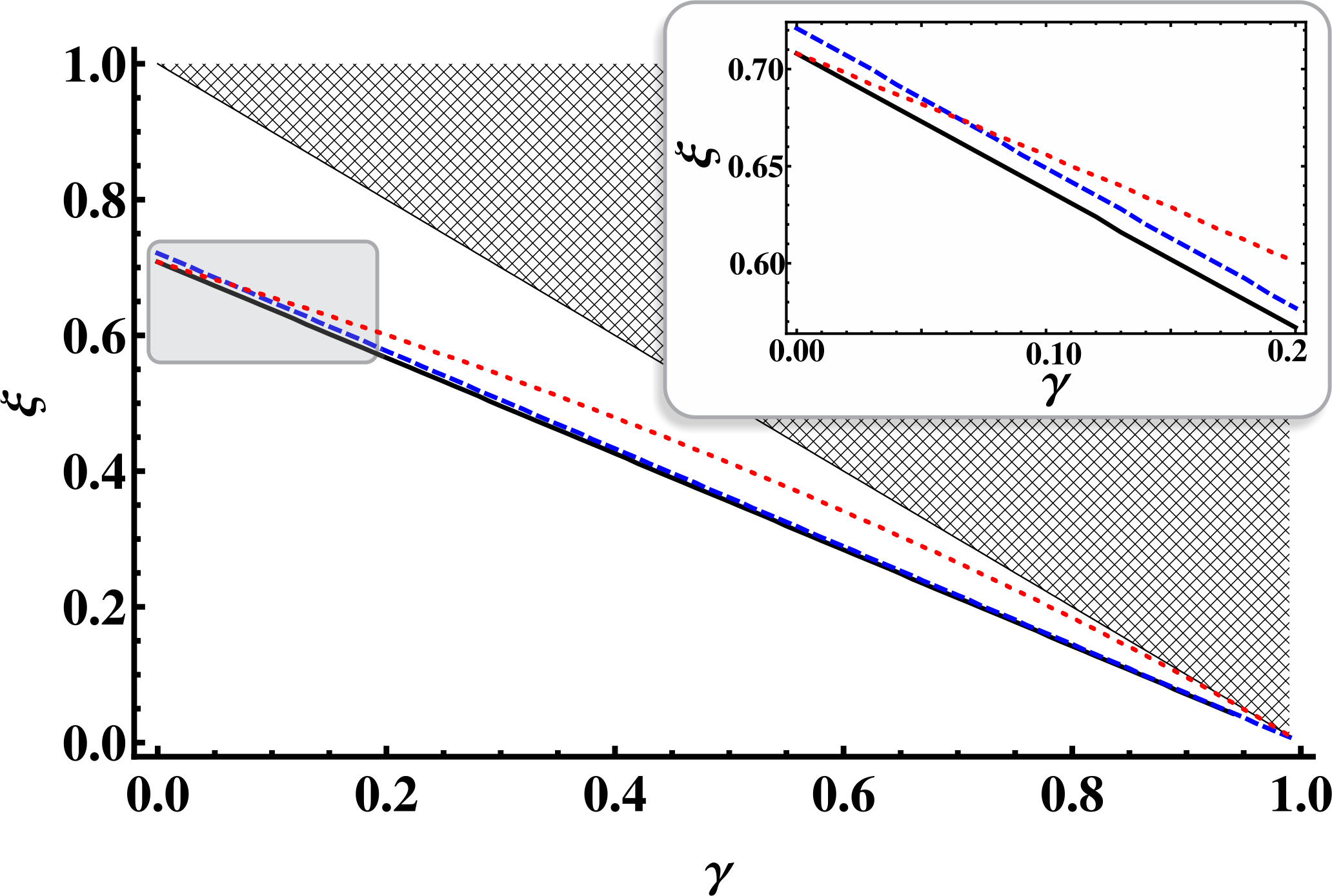}
\caption{IC vs LO performance. Comparison of both principles for detecting supra-quantum
correlations in the (2,2,2) scenario. We consider the family of
correlations parametrized as $P(ab|xy)=\xi PR(ab|xy) + \gamma
P_L(ab|xy) + (1-\xi - \gamma) P_\mathbb{I}$, where
$P_L(ab|xy)=\delta_{a,0}\delta_{b,0}$. The curves show the bounds
provided by the 1+AB level of the NPA
hierarchy~\cite{navascues2007,navascues2008} (black),  LO$^2$ (blue
dashed), IC (red dotted), and the edge of the crossed-out region
corresponds to NS correlations and bounds the allowed parameter
space. Note that when $\gamma \rightarrow 0$ (see inset), IC
approximates the quantum set better than  LO$^2$, which is consistent
with the fact that IC recovers Tsirelson's bound for $\gamma=0$,
while LO reaches $\approx 0.72$  for two copies of the device.  However, LO beats the known IC bound for
other parameter values, ruling out correlations that were not excluded
before.}\label{fam}
\end{figure}

It is also interesting to compare LO with information causality
(IC)~\cite{Pawlowski2009a}, another proposal for a physical
principle to single out quantum correlations. A natural question
then is if and when LO can do better than IC in ruling out
supra-quantum correlations. Following Allcock \etal\cite{Allcock},
we study LO predictions for different families of no-signaling
correlations. In some situations LO provides a bound to the set of
quantum correlations which is tighter than the known bounds
obtained from applications of IC, as can be seen in \figref{fam}.
Hence, LO rules out correlations that were not excluded before by
IC.

So far we have focused on how some bipartite supra-quantum
correlations can be excluded by LO. Due to the intrinsically
multipartite formulation of the principle, we expect it to be of
particular relevance for the study of genuine multipartite
correlations among more than two parties. As an example, we study
how well LO performs for  extremal no-signaling nonlocal
correlations in the $(3,2,2)$ scenario. All these extremal NS
boxes were computed in Ref. \cite{tri-box} and any NS correlations
in this scenario can be obtained by mixing them. These extremal
boxes can be grouped into 46 equivalence classes under symmetries,
the first class corresponding to deterministic local points, while
the other 45 are non-local.  The latter can be interpreted as the
maximally non-local correlations in the $(3,2,2)$ scenario
compatible with the NS principle. LO can be shown to rule out all
these maximally non-local boxes. We follow a method similar to the
one for PR-boxes presented in Methods, although for some cases the
tripartite boxes required a slightly more general approach in
terms of weighted graphs.
 The details of these derivations are beyond the scope of
the present manuscript and will be presented elsewhere.
We find that every maximally
non-local box in $(3,2,2)$ violates either LO$^1$ or LO$^2$, and
hence cannot have a quantum realization.

We conclude this part with a couple of observations. First, LO is,
to our knowledge, the first principle able to rule out all the
know examples of extremal non-local boxes. We have shown it first
for the only maximally non-local box in the $(2,2,2)$ scenario
(the PR-box), then for maximally non-local boxes in the $(2,m,2)$
and $(2,2,d)$ scenarios, as well as for all extremal boxes in the
$(3,2,2)$ case. Second, the intrinsically multipartite formulation
of LO allows detecting correlations for which any bipartite
principle fails. For instance, the correlations provided
in~\cite{gallego} violate GYNI and, consequently, also LO, but
satisfy every bipartite principle. Also box number 4
in~\cite{tri-box} is an example of a tripartite no-signaling box
which cannot be ruled out by any bipartite
principle~\cite{3box-IC}, but which violates LO.

\section*{Discussion}

Multipartite principles are necessary to understand the structure
of quantum correlations from an information perspective. Local
orthogonality is a principle for correlations that has an
intrinsically multipartite formulation. It has a natural
interpretation in terms of distributed guessing problems.
Moreover, it can be used in the study of correlations by applying
standard techniques from graph theory. We have shown here that the
principle gets very close to the set of
quantum correlations and detects all known
 maximally non-local boxes as supra-quantum, including
those for which any bipartite principle fails.

We believe that LO will be an essential ingredient for the
characterization and understanding of correlations. An intriguing
question is to understand the exact relation between this set and
the set of quantum correlations.  Miguel Navascu\'es first noticed
that the set of quantum correlations is strictly smaller than
$\mathcal{LO}^\infty$, as also proved in \cite{FLS}. Are there simple
additional principles that, together with LO, completely
characterize the set of quantum correlations?

Another interesting line of research consists in extending our
approach, here formulated for Bell scenarios, to other frameworks.
In fact, the principle is rather general: after introducing a
notion of orthogonality in a space of events, the whole machinery
of graph theory and copies of the graph (networks) automatically
applies.  A natural framework for its application is the
study of contextuality scenarios where, as mentioned, there is a
natural notion of orthogonality (or exclusiveness). In fact, our
approach has successfully been applied to the study of quantum
contextuality in several recent
works~\cite{FLS, cabello, Yan13}.

\section*{Acknowledgements}
Discussions with A. Cabello, M. Navascu\'es and M. Paw\l{}owski
are gratefully acknowledged. Research at Perimeter Institute is
supported by the Government of Canada through Industry Canada and
by the Province of Ontario through the Ministry of Economic
Development and Innovation. A.L. was supported by the SNF through
the National Centre of Competence in Research ``Quantum Science
and Technology''. R.A. acknowledges the support from EU Project
AQUTE and Spanish MINCIN through the Juan de la Cierva program.
R.C. was supported by the German Science Foundation (grant CH
843/21) and the Excellence Initiative of the German Federal and
State Governments (grant ZUK 43). A.B.S., J.B.B. and A.A. were
supported by the European PERCENT ERC Starting Grant, Q-Essence,
QCS and DIQIP Project, by the Spanish project FIS2010-14830, 
the AP2009-1174 FPU PhD grant, and CatalunyaCaixa.

\appendix

\section*{Methods}

\noindent\textbf{Maximally
difficult DGPs and LO inequalities.} Our goal here is to prove that
imposing that correlations do not provide any advantage for DGP
involving maximally difficult functions $f$ is equivalent to LO.
For such an $f$ and any correlations $P(a_1\ldots
a_n|x_1\ldots x_n)$, providing no advantage for the DGP defined by $f$ means that
\begin{equation}
\label{dgpineq}
\frac{1}{|S|} \sum_{(a_1,\ldots,a_n)\in S} P \left( a_1\ldots a_n
| f_1(a_1,\ldots,a_n) \ldots f_n(a_1,\ldots,a_n) \right) \leq
\frac{1}{|S|},
\end{equation}
where $f_1, \ldots, f_n$ refer to the components of the vector $f$,  and $x_j = f_j(a_1,\ldots,a_n)$ is the input that party $j$ receives.
Note that, for simplicity, and since the goal of the parties is to
provide a correct guess of the initial parameters, we slightly
abuse notation and replace all $\widetilde a_j$ by $a_j$. In
order to prove the correspondence, we now show that functions $f$
that are maximally difficult for classical players are precisely
those which have the property that if $f(a_1,\ldots,a_n)$ and
$f(a'_1,\ldots,a'_n)$ are both defined, then there exists some $j$
for which $a_j \neq a'_j$, but $f_j(a_1,\ldots,a_n) =
f_j(a'_1,\ldots,a'_n)$. Given that $f$ varies over all those
partial functions having this property, the DGP
inequalities~(\ref{dgpineq}) define all LO inequalities in the $(n,m,d)$ scenario.

We first prove the 'only if' direction by contradiction. Assume
there exist $a_1,\ldots,a_n$ and $a'_1,\ldots,a'_n$, both on which
$f$ is defined, such that for every party $j$, either $a_j = a'_j$
or $f_j(a_1,\ldots,a_n) \neq f_j(a'_1,\ldots,a'_n)$ holds true.
Then the following classical strategy performs better than random
guessing: for those $j$ with $a_j=a'_j$, let them output this
particular value independently of their input; for those with $a_j
\neq a'_j$ and $f_j(a_1,\ldots,a_n) \neq f_j(a'_1,\ldots,a'_n)$,
choose some function $g_j$ such that $a_j =
g_j(f_j(a_1,\ldots,a_n))$ and $a'_j = g_j(f_j(a'_1,\ldots,a'_n))$,
and let them output $g_j(x_j)$. This strategy recovers all correct
values both for the $a_1,\ldots,a_n$ as well as for the
$a'_1,\ldots,a'_n$ and therefore performs better than random
guessing.

Conversely, we need to show that if $f$ has this property, then
using local operations only cannot be more successful than random
guessing. Thanks to convexity, it is enough to consider
deterministic local strategies. If a deterministic strategy is
better than random guessing, there needs to exist at least
$a_1,\ldots,a_n$ and $a'_1,\ldots,a'_n$ such that the strategy
works on both of these. In particular, this means that, for each
party $j$, either $a_j = a'_j$, or party $j$ needs to be able to
tell the two cases apart via $x_j$, so that $f_j(a_1,\ldots,a_n)
\neq f_j(a'_1,\ldots,a_n)$. This implies that $f$ cannot have the
property described above.\\

\noindent\textbf{NS and  LO$^1$ in bipartite scenarios.} We prove that in the bipartite case, LO${}^1$ and NS
define the same set of correlations. Although this is already
known~\cite{csw}, here we give a slightly different proof which
emphasizes the connection with LO. To simplify the notation, in
the bipartite scenario, measurements and results by the two parties
are labeled by $x,y$ and $a,b$, so that correlations read
$P(ab|xy)$.

Let us start by characterizing the possible sets of locally
orthogonal events. Recall that two events are locally orthogonal
if for at least one party the settings are identical but the
outcomes are different. Consider a set of locally orthogonal
events which contains $(ab|xy)$ and $(a'b'|x'y)$ with $x'\neq x$.
Then, this set cannot contain any event of the form
$(a''b''|x''y')$ with $y' \neq y$, because it could not be locally
orthogonal to both other events. From this intuition, we find that
the sets of pairwise orthogonal events are either
$\{(ab|x\omega_A(a)):a,b=0,\ldots, d-1 \}$ for fixed $x$, or
$\{(ab|\omega_B(b)y):a,b=0,\ldots, d-1 \}$ for fixed $y$ with
$\omega_{W}: \{0,\ldots,d-1\} \longrightarrow \{0,\ldots,m-1\}$
$(W=A,B)$ being some map.

We start by showing that sets of the first kind have the desired
property; the proof for sets of the second kind is analogous. Take
two such events $(ab|x\omega_A(a)) \neq (a'b'|x\omega_A(a'))$. Then
either $a\neq a'$ and orthogonality holds on Alice's side, or
$b\neq b'$ and orthogonality follows from Bob. This proves that we
have a set of LO events. To see that the set is maximal, consider
an arbitrary event
$(\widetilde{a}\widetilde{b}|\widetilde{x}\widetilde{y})$. If
$\widetilde{x}=x$ and $\widetilde{y}=\omega_A(\widetilde{a})$,
then this event is already in the set. Otherwise, LO fails between
$(\widetilde{a}\widetilde{b}|\widetilde{x}\widetilde{y})$ and
$(\widetilde{a}\widetilde{b}|x\omega_A(\widetilde{a}))$. Hence it
is impossible to add any event to the set, i.e. the set is
maximal.

Now we prove that every maximal LO${}^1$ set is of one of these two
forms. It is enough to show that every LO${}^1$ set is contained in a
set of this form. As noted above, for events in an LO set, one of
the parties is restricted to using a single input. Hence, without
loss of generality, we can take $x$ to be fixed. Since every two
orthogonal events differ on at least one outcome, there exists a
function $\omega(a,b)$ such that every element in the set is of
the form $(ab|x\omega(a,b))$. We complete the proof by showing
that $\omega(a,b)$ does not depend on $b$. The existence of $a$
and $b$, $b'$ with $\omega(a,b)\neq \omega(a,b')$ would imply that
$(ab|x\omega(a,b))$ and $(ab'|x\omega(a,b'))$ are not orthogonal,
contradicting the assumption.

We denote by $\mathcal{NS}$ the set of distributions $P(ab|xy)$
satisfying no-signaling. The conditions for no-signaling are
\begin{equation}
\label{ns}
\sum_{b=0}^{d-1} P(ab|xy) = \sum_{b=0}^{d-1} P(ab|xy') \,, \quad
\sum_{a=0}^{d-1} P(ab|xy) = \sum_{a=0}^{d-1} P(ab|x'y) ,
\end{equation}
and we also have the normalization conditions $\sum_{a,b=0}^{d-1}
P(ab|xy)=1$, $\forall x,y$. We now prove that $\mathcal{LO}^1 =
\mathcal{NS}$, in the present bipartite setting.

{\bf $\mathcal{LO}^1 \subseteq \mathcal{NS}$:} all optimal LO${}^1$
inequalities are of the form $\sum_{a,b=0}^{d-1}
P(ab|x\omega_A(a))\leq 1$, modulo exchanging the parties. We fix
any $a_0$, $y$ and $y'$ and consider the function $\omega_A(a)= y$
if $a=a_0$ and $\omega_A(a)= y'$ if $a\neq a_0$. The LO inequality
yields $\sum_{a\neq a_0,b=0}^{d-1} P(ab|xy')+\sum_{b=0}^{d-1}
P(a_0b|xy)\leq 1$. Together with the normalization equation
$\sum_{a,b=0}^{d-1} P(ab|xy')=1$, this implies $\label{LO2NS}
\sum_{b=0}^{d-1} P(a_0b|xy) \leq \sum_{b=0}^{d-1} P(a_0 b|xy')$.
Since the same inequality can be derived with $y$ and $y'$
interchanged, we find that it actually needs to be an equality,
which is~(\ref{ns}).

{\bf $\mathcal{NS} \subseteq \mathcal{LO}^1$:} start from the
normalization condition $\sum_{a,b=0}^{d-1} P(ab|xy)=1$. Using the
no-signaling equations~(\ref{ns}), we can transform it into an
equality of the form $\sum_{a,b=0}^{d-1} P(ab|x\omega_A(y))=1$ for
any given $x$ and $\omega_A$. It follows that $\mathcal{NS}
\subseteq \mathcal{LO}^1$, and thus $\mathcal{LO}^1 = \mathcal{NS}$ in the bipartite scenario.\\

\noindent\textbf{Quantum correlations satisfy LO.} It is straightforward to see that
LO inequalities are satisfied by the set of quantum correlations,
denoted $\mathcal{Q}$, that is, $\mathcal{Q} \subseteq
\mathcal{LO}^\infty$. For simplicity, we give the proof for an inequality
involving two LO events. The generalization to an arbitrary LO
inequality, and in particular to any number of copies, will be presented elsewhere, and follows from the property that in quantum mechanics, a set of
projective measurements which can be pairwise implemented, can be jointly implemented (i.e. pairwise orthogonality between projectors implies orthogonality of all the projectors).

Consider two LO events $e_1=(a_1 \ldots a_n|x_1 \ldots x_n)$ and
$e_2=(a'_1 \ldots a'_n|x'_1 \ldots x'_n)$ with $a_i \neq a_i'$,
$x_i=x_i'$, and the corresponding inequality $p\,(e_1)+p\,(e_2)\leq 1$.
The maximization of the sum of these two probabilities over quantum correlations reads
\begin{equation}\label{qlo}
    \max_{\ket{\Psi},\{\Pi_{a_i}^{x_i}\}}
    \bra{\Psi}(\Pi_{a_1}^{x_1}\otimes\ldots\otimes\Pi_{a_i}^{x_i}
    \otimes\ldots\otimes\Pi_{a_n}^{x_n}+\Pi_{a'_1}^{x'_1}\otimes
    \ldots\otimes\Pi_{a_i'}^{x_i}\otimes\ldots\otimes
    \Pi_{a'_n}^{x'_n})\ket{\Psi} ,
\end{equation}
where the maximization runs over all possible states $\ket{\Psi}$ and projectors $\{\Pi_{a_i}^{x_i}\}$ acting on an arbitrary Hilbert space. Note that the maximization can be done over projective measurements without loss of generality, as the ancilla needed for a general measurement can be absorbed in the definition of the state $\ket{\Psi}$. The term in the parenthesis is equal to the sum of two orthogonal projectors, as $\Pi_{a_i}^{x_i}\Pi_{a_i'}^{x_i}=0$. Thus, this sum is
upper bounded by the identity operator and the LO inequality follows.\\

\noindent\textbf{Two PR-boxes violate LO}. We classify events as either ``possible'' or ``not possible'' for this many-copy box as follows. An event $(ab|xy)$ for one PR-box is possible if $a\oplus b = xy$. Hence, PR-box correlations may be written as
\begin{equation}\label{PR-pos}
PR(ab|xy) = \begin{cases} \frac{1}{2} & \text{if the event is
possible}, \\ 0 & \text{otherwise}. \end{cases}
\end{equation}
In the case of $k$ boxes, an event $(a_1 b_1 \cdots a_k b_k |
x_1y_1 \cdots x_k y_k )$ is possible iff $a_j \oplus
b_j=x_j y_j$ for all $j \in \left\{ 1, \ldots, k\right\}$.
Then, the general form for the $k$-box probability (\ref{def-NPR}) is
\begin{equation}
\label{manyPR}
P(a_1 b_1 \cdots a_k b_k | x_1y_1 \cdots x_k y_k ) =
\begin{cases} 2^{-k} & \text{if the event is possible}, \\ 0  & \text{otherwise}.\end{cases}
\end{equation}

\begin{figure}
\centering
\includegraphics[scale=0.3]{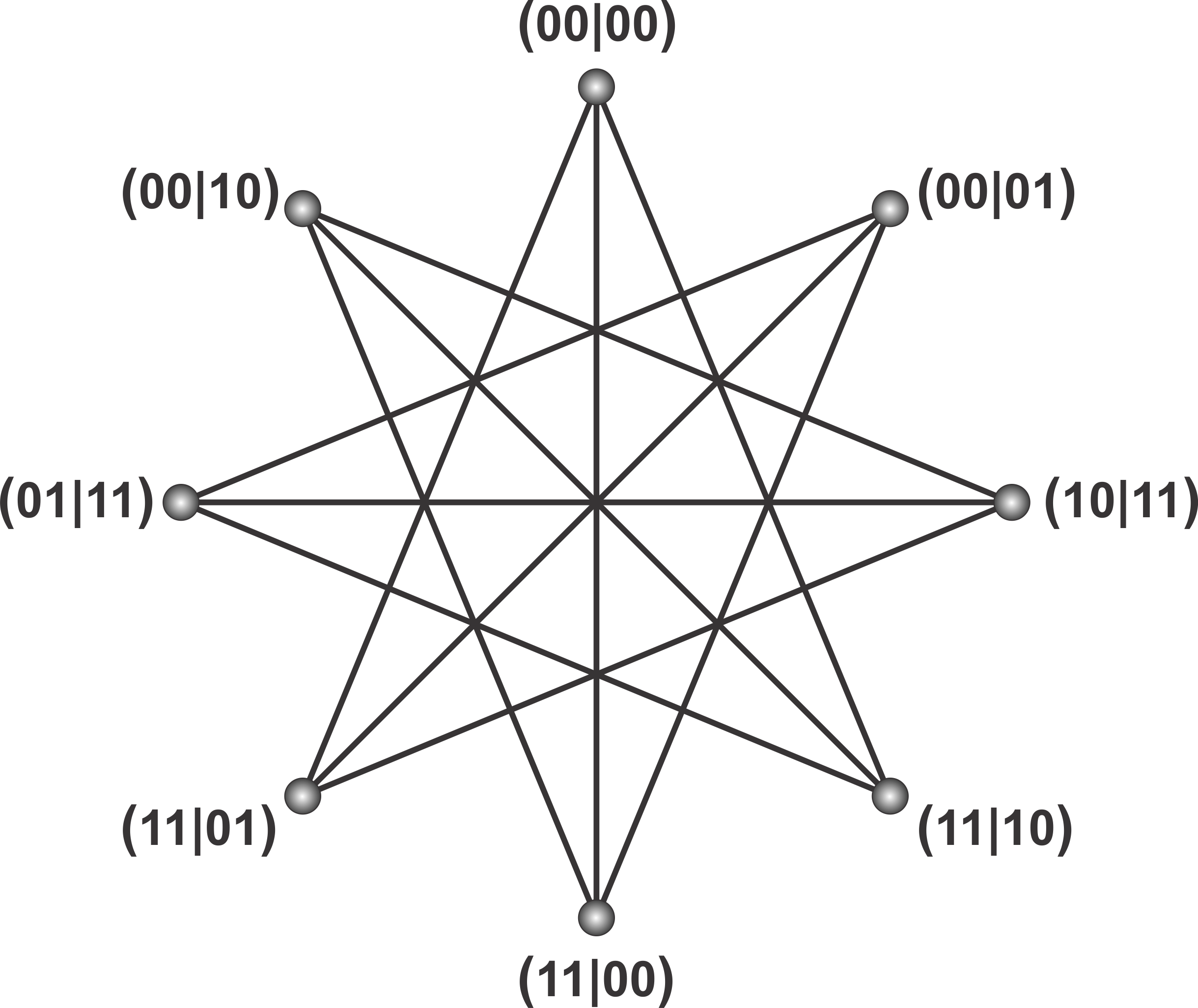}
\caption{Orthogonality graph of possible events for a single
PR-box. It coincides with Fig.~2 in \cite{SBBC}, where the authors
study the CHSH inequality. }\label{grafoPR}
\end{figure}

Consider a clique $C \subseteq V$ in the orthogonality graph
$G=(V,E)$ of the scenario $(2k,2,2)$ and the corresponding LO
inequality $LO(C)$. Define the set $C_p\subseteq C$ to be the
subset of possible events in $C$. Then, the multipartite
box~(\ref{manyPR}) violates $LO(C)$ if, and only if, it violates
$LO(C_p)$. In particular, in order to exclude the PR-box, it is
sufficient  to find a clique of size larger than $2^k$ in the
orthogonality graph $G_\mathrm{poss}=(V_p,E_p)$ of possible events
for box~(\ref{manyPR}). This problem becomes significantly easier,
since $|V_p|=8^{k}$, compared to $|V|=16^{k}$ for the initial
graph (compare Figs.~\ref{OG2} and~\ref{grafoPR}). Already for
$k=2$, there exist cliques of size larger that $4$. We found that
all of them have size $5$, and one example is given by:
$\{ (0000 | 0000) ,\: (1110 | 0011) ,\: (0011 | 0110) ,\: (1101 |
1011) ,\: (0111 | 1101)\}.$

Consider now 2 copies of a noisy PR-box given by $P_{q} = qPR + (1-q)
P_{\mathbb{I}}$, where $P_\mathbb{I} (a,b|x,y)=1/4$ for
all $a,b,x,y$. Now all events become possible and one should
consider the full list of LO inequalities for $(4,2,2)$. However,
one can still consider this five-term inequality as associated to
a non-maximal clique in the orthogonality graph of the $(4,2,2)$
scenario, and complete it to a maximal clique. This gives an
inequality with additional terms corresponding to events that are
impossible for the PR-box, but which are possible for a noisy
PR-box. In this way, we have found that the distribution
\begin{equation}
P(a_1 b_1 a_2 b_2|x_1y_1x_2y_2) = \left( q PR(a_1b_1|x_1y_1)
+ (1-q)\tfrac{1}{4}\right) \cdot \left( q
PR(a_2b_2|x_2y_2) + (1-q)\tfrac{1}{4}\right)
\end{equation}
violates LO for $q \gtrsim0.72$. An example of such an LO
inequality is given by the following set of ten LO events:
$\{(1111 | 0000),(1100 | 1010),(0100 | 1100),(0011 | 0001),(0010 |
0111),(1011 | 0000),\\(0101 | 1100),(1101 | 1100),(1010 |
0110),(1001 | 0100) \}.$

It appears plausible to conjecture that the generalization of the
previous approach to an arbitrary number of parties converges to
Tsirelson's bound $q=1/\sqrt{2}\approx 0.707$ in the limit of an
infinite number of parties, although we did not yet find any
proof.

\bibliographystyle{naturemag}

\end{document}